\documentclass[sn-mathphys,Numbered]{sn-jnl}

\usepackage{graphicx}%
\usepackage{multirow}%
\usepackage{amsmath,amssymb,amsfonts}%
\usepackage{amsthm}%
\usepackage{mathrsfs}%
\usepackage[title]{appendix}%
\usepackage{xcolor}%
\usepackage{textcomp}%
\usepackage{manyfoot}%
\usepackage{booktabs}%
\usepackage{algpseudocode}%
\usepackage{listings}%
\usepackage{caption} 
\usepackage{subcaption}
\usepackage{amsmath,amsfonts}
\usepackage{amsthm,amssymb}

\theoremstyle{thmstyleone}%
\theoremstyle{thmstyletwo}%

\theoremstyle{thmstylethree}%

\begin{document}

\title[ ]{The Approach of Sliced Inference in Systems of Stochastic Differential Equations with   Comments on the Heston Model}


\author*[1 ]{\fnm{Ahmet Umur} \sur{\"Ozsoy}}\email{umurozsoy@gmail.com}

\affil*[1]{\orgdiv{Department of Industrial Engineering},  \orgname{Gebze Technical University}, \orgaddress{  \city{Kocaeli}, \postcode{41400},  \country{Turkey}}}



\abstract{Stochastic  differential equations have been an important tool in modeling complex financial relations, equipped with the possibility of being multidimensional to better oversee complexities inherent in finance. This multidimensionality, however, comes with a  larger parameter space to estimate. Therefore, via  a dimension reduction method, Sliced Inverse Regression, we aim to reduce high-dimensional parameter space to a reduced feature space and aim to estimate the parameters on this new featured space rather than using full data structure to lower computational costs. For this study, we closely study the Heston model, and remark our methodology of inference on this chosen model. }

\keywords{dimension reduction, parameter inference, volatility modeling, maximum likelihood}	




\maketitle

\section{Introduction}\label{sec1}

Living in an era of increased computational capabilities and abundance of data are both incredibly advantageous and disadvantageous.
On one side, these realities motivate and challenge us all to explore further complex relationships among variables, yet such endeavors require stronger than before computational capabilities and higher volume of data.
This, rather in fact, constitutes the new foundations of higher computational capabilities, that will fall short soon considering the fact that much higher computational capabilities will be required.
Although we welcome such vicious circles enlarging our comprehension, it is also paradoxical not to pose a simple question whether or not we could achieve some results effectively without resorting to more complex techniques or requesting more data.
This simple questions is highly straight forward.
We ask whether we could achieve the same (or acceptable) results by employing simpler (i.e. less computationally intensive)  techniques for the purpose of parameter estimation, and perhaps in a much faster way. 
This brings us to the concept of dimensionality reduction and its possible usage in inference of stochastic differential equations; especially in multidimensional ones.
Dimensionality reduction is a fundamental technique formulated with a simplistic expectation.
The expectation mentioned is to preserve the \emph{meaning} of data while reducing the number of the input space.

The number of the input space, here we refer to, is the high-dimensional perspective of the model at hand, such as the Heston model. 
Dealing with such high-dimensionality could lead to   highly increased complexity in computational matters and, above all higher  required time for computation. 
Although we could give examples that might be computationally intensive examples of such methods, for instance applying t-SNE to large data sets could be an interesting example, \cite{van2008visualizing}.  
Therefore, we remark that expecting enhanced computational efficiency via dimensionality reduction could be possible given that such methods, in all its basic essence, delivers a lower-dimensional data. 

The degree of how well the original feature of a data set is preserved is also related to the degree of information loss incurred after dimensionality reductions methods. 
Especially if the variability captured by the dimensions reduced is significantly lower than acceptable levels, then applying such methods might not be feasible at all. 
We also need to mention that the levels of acceptability is highly subjective, and drawing conclusions on such bounds is no easy way. 
Another note on this, choosing the appropriate methods could also be challenging as subject-related expertise might be strongly needed. 
From this perspective, we remark on our study we select a bit generalized method coined \emph{Sliced Inverse Regression} in \cite{li1991sliced} for reasons we now explain.

Given our selection of the Heston model, throughout this study we approach the concept of inference from the perspective of financial modeling.
On top of the complexity of inference, in such models, the issue of interpretation exists.
This signifies the fact that the method chosen for dimensionality reduction should not severely affect the interpretative nature of the modeling; unfortunately we are unaware of such studies conducted.
As we stress, in dealing with financial models, one should exercise as much caution as possible. 
For instance, a well-known one of this kind, Principal Component Analysis (PCA) could be considered as unsupervised; meaning that  problems of interpretation might incur in modeling relationships. 
The reason is quite simple as PCA is designed to find variability in predictors with no regard to the target variable
Sliced Inverse Regression (SIR), on the other hand, provides a supervised learning environment that could be highly valuable in predictive analyses.
What is more interesting is that SIR is computationally very direct, and could be considered an effective data preprocessing tool in numerous tasks of estimation.
Therefore, its simplistic nature makes it a great candidate from a financial perspective.

Another major feature of SIR is that it could deal with large samples quite efficiently; for instance \cite{hsing1992asymptotic}. 
As large-scale simulated data has been a standard practice in option pricing, risk management, portfolio optimization and so on.
With this tradition of high volume of synthetic \emph{test} data along with asymptotic features of SIR align well with each other.
SIR, perhaps most importantly, neither requires nor impose a certain parametric form between predictors and the target variable (or alternatively, independent variables and dependent variable with the analogy of statistics). 
To reframe it with a newer terminology introduced mainly in the last decade; SIR could go \emph{model-free}.
Therefore, it could be employed in inference of wide range of  stochastic differential equations.
 
Although, by definition, SIR is a linear transformation in essence; it might achieve linear approximation of nonlinear relationships inherent in some models.
This further contributes to the fact that SIR does not make overly restrictive assumptions about the structure of data.
But above all, given the dimensions reduced; a modeler should be willing to accept certain amount of error in inference as the loss of information is expected in the procedure of SIR, or any of its kind.

By kind, though, we refrain ourselves from categorizing SIR other than treating it as a linear transformation, by generalization provided in \cite{cunningham2015linear}, SIR is classified into \emph{sufficient dimensionality reduction} of dimension reduction methods.
Furthermore, in  \cite{adragni2009sufficient}, SIR is labeled as \emph{moment-based inverse reduction} posing the linearity condition and constant covariance condition for both we hold largely nonrestrictive views in our study. 
For more technical details, we refer to \cite{adragni2009sufficient} in which an engaging discussion of SIR along with its weaknesses could also be found. 
Before we move on the details of our alternative approach; we find it better to mention inference in stochastic differential equations; without the dimensionality reduction; hence (we call) \emph{direct inference}.

Models incorporating stochastic differential equations  with the intent to model complex relationship or structures are not new in the literature.
Ever since the publication of the famous Black-Scholes-Merton model; stochastic differential equations with more dimensions as well as higher complexities have arisen in the literature, and are likely increase both by diversity and complexity.
Even a study published in 1990, \cite{sharp1990stochastic}, delivers interesting perspectives of such models and the progress achieved so far back then.
Although there has been a wave of such models surfacing for decades; there has also been a strong necessity to fit such models to real or synthetic data.
With the increased complexity, at the same time, challenges in inference are likely.
We could observe that a study, that could be found in \cite{nielsen2000parameter}, covering the progress of research on parameter estimation methods for stochastic differential equations during between 1981 and 1999 sheds further light to this reality.
We also refer to a more recent review given in \cite{donnet2013review} as the study categorizes methods structurally; and some challenges are discussed thoroughly.

The review given in \cite{craigmile2023statistical} perhaps is one of the most complete and friendliest introduction out there in which well-known examples such as the Ornstein-Uhlenbeck, Black-Scholes-Merton  (geometric Brownian motion) and Cox-Ingersoll-Ross processes are introduced along with their closed-form transition densities.
Along with exact transition densities of such models, the idea of approximating transition densities is covered briefly and nicely as well.
The review also draws attention to some properties of Maximum Likelihood Estimation on one-dimensional stochastic differential equations.
We particularly refer to the references therein \cite{craigmile2023statistical}
Another notable survey in addition to studies in \cite{nielsen2000parameter, donnet2013review} could be found in   \cite{sorensen2004parametric}.   
A highly numerical review is given in \cite{brouste2014yuima} along with brief properties of numerous stochastic differential equations.

Interestingly enough; there are two  that we could consider as  thorough resources when it comes to inference on such processes; namely \cite{iacus2008simulation} and a later version of the same author covering some further concepts in \cite{iacus2018simulation}. 
We personally believe the earlier version, \cite{iacus2008simulation}, could appeal to especially to market practitioners in financial modeling.
The mentioned books, we strongly refer to, cover almost all aspects of inference on stochastic differential equations; although the emphasis is mainly on the one-dimensional ones. 
The fact that stochastic differential equations are used by many fields such as finance, biology, physics etc. contributes to our view on being restrictive on citing (or comparing) many of the available methods in comparison of effectiveness with each other. 
To remain concise and focused, we limit detailed comparisons and conclude this section with selective references, moving on to our own  suggested framework of inference.

With both current standing of inference and our reasons to opt for SIR briefly explained; we believe we are ready to define our purpose and research question. 
What we call direct inference has already rigorous and effective approaches in estimating parameters of such models.
However, such models with a larger input (parameter) space tend to create difficulty in matters of inference; especially in  models with several sources of uncertainty like the Heston model, \cite{heston1993closed}.
The question is the applicability and effectiveness of SIR as a data preprocessing step first in estimating the parameters of stochastic differential equations, hence (we call) \emph{sliced inference}.
In our set-up; we keep the method of estimation the same between direct inference and sliced inference; the Maximum Likelihood Estimation (MLE);
yet only the ordering is no longer the same.

Formally, SIR could stand an interesting and effective candidate for a distinct step of preprocessing in inference of such models.
Its design with the principles of simplicity and its model-free structure add up to our argument in support of SIR. 
Consider the functional below with $p$ and $k$ some positive numbers such that $k < p$
\begin{equation}\label{eq:funct}
	Y = f (\beta_1^\top x, \ldots, \beta_B^\top x, \epsilon  ),
\end{equation}
implying that the target variable $Y$ is given (or could be given) by a $p$-dimensional independent variable $x$ via the reduced $B$ dimensional vector, i.e. $[ \beta_1^\top x, \ldots, \beta_B^\top x  ]^\top$.
SIR, by definition, aims to find the (sub-)space spanned by $\beta$ vector; which are termed \emph{effective dimension reduction space}.
What increases our interest in SIR is as put forward by \cite{chen1998can}; there are no assumptions in regard to the distribution of the error, $\epsilon$, and $f$. 
This particular aspect is especially welcome as it's not always possible or plausible to assume the true probability distribution.

We also bring up an important aspect assumed in SIR, linearity of transformation. 
An intriguing anecdote reported by \cite{cook1991sliced} states that the SIR is usually not overly sensitive to the linearity condition.
This, in a way, prompts us to investigate the potential of the SIR in models (e.g. the Heston model) where non-linearity is likely and often inherent with high degree.  
Modeling through log-returns is a common practice in finance along with discretization of stochastic differential equations.
Another major feature suggested by \cite{chen1998can} is that there no need in logarithmic transformation on $Y$.
Although there is a vast amount of studies in the literature based on log-return view of modeling, in applying SIR this is not necessarily advantageous.
For instance, suppose that a discretization scheme yields $S_{t+1} = S_t + \mu S_t \Delta t + \cdots $, simple division of both sides by $S_t$ 
would suffice as an initial modeling and estimation point without ever resorting to the exact solutions involving logarithmic returns.
We also take this view to test whether or not \cite{chen1998can} accurately points this out.

Consider a testing environment in which models are rigorously fitted to data and backtested for reasons of model validation or portfolio construction; or any field in need of sufficiently large amount of simulation or paths generated by a stochastic differential equation.
To present a better idea; cross validating models or stress testing require high amount of data; then our question becomes whether or not we can employ SIR before fitting to the data and still get viable results.

The aim is to apply SIR on a model with a large number of covariates such that further estimation procedures  (MLE)  might lead to less computational time, considering the loss of information too of course.
With this aim, we simulate data via the Heston model with fixed randomness for both cases. 
Afterwards; we carry out direct inference and sliced inference to see if the loss of information (lost predictive performance) is acceptable given the gain of computational time. 
Supposedly, if SIR manages to reduce the dimensionality, the low-dimension obtained might simplify how parameter dependence is structured which might lead to acceptable estimates of the parameter space. 

To sum up; in this study we aim to investigate the ability of SIR as a data preprocessing tool before inference on (financial) models built on stochastic differential equations
While aiming for accuracy, our major interest lies in the computational efficiency.
Surely, given the loss of information incurred after applying SIR; the direct inference should show better performance in comparison to the sliced inference.
However, the performance alone is a difficult measure to define excluding the required time for computation.
Yet we choose to leave the trade-off between such metrics to practitioners.

The study consists of three more sections.
In the next one, we develop our alternative approach   providing as necessary as possible detail while assuming a style of keeping further theoretical perspective at bay.
Later on we provide numerical illustrations and conclusions suggesting further possible extension.
To the best of our knowledge, our study is the first one in its unique approach to inference of models built on stochastic differential equations. 
Lastly; in our study, we use the terms variability, variation or explainability interchangeably whereas we reserve the term variance for a statistical measure of dispersion.

\section{The approach in detail}

In this part, we provide the details of our suggested approach in estimation of parameters.
In the first part, we introduce the Heston model briefly, and provide details of the associated log-likelihood function following \cite{dunn2014estimating} closely. 
Although there are various approaches of inference on the Heston model such as \cite{ait2007maximum} or \cite{wang2018parameter}; we prefer to follow \cite{dunn2014estimating} for reasons we explain shortly.
In the second part, we introduce \emph{Sliced Inverse Regression} first; and our approach with detail.
Given that we   have a reduced data after SIR, we also derive the reduced log-likelihood required to extract the parameters.
In both cased, our method of estimation is MLE.
While we acknowledge there are alternatives, we prefer a simplistic tool in our manuscript.

\subsection{Direct inference }

The Heston model is a two-dimensional system of stochastic differential equations with one dimension on asset process and the other on variance process, \cite{heston1993closed}. 
For derivation and implementation of our alternative estimation procedure we select the Heston model; and there exist several reason behind it.
The Heston model incorporates a stochastic volatility component, making the asset price more realistic with fat-tails and volatility clustering.
Besides, given our purpose on dimensionality reduction; the Heston model is packed with numerous parameters.
Lastly, the parameter estimation is a challenge on its own.

Given the parameters, $\mu$  as the drift of the asset price, $v_t$ variance ,$\kappa$ as the rate of mean reversion, $\theta$ as long-run variance level and $\sigma$ as volatility of volatility; the Heston model is defined such as
\begin{align} \label{eq:Heston_sde_original}
	dS_t  &= \mu S_t dt + \sqrt{v_t} S_t dW_{1,t}, \\ 
	dv_t &= \kappa ( \theta - v_t) dt + \sigma \sqrt{v_t} dW_{2,t},
\end{align}
where $dW_{1,t}$ and $dW_{2,t}$ are correlated standard Brownian motions, and the degree  of the correlation  is quantified by $\rho$.
Applying Euler-Maruyama discretization; we obtain
\begin{align}
	S_{t+1} &= S_t + \mu S_t dt + \sqrt{v_t} S_t \sqrt{dt}Z_s \\ \label{eq: s_disc},
	v_{t+1}&=v_t+  \kappa(\theta-v_t)+\sigma \sqrt{v_t} \sqrt{dt} Z_v,
\end{align}
where $Z_s, Z_v \sim \mathcal{N}(0,1) $. In the original version laid out in \eqref{eq:Heston_sde_original}, there are two Brownian motions embedded with a correlation, denoted by $\rho$. 
To incorporate the inherent correlation, we define two independent standard normal variables $Z_1, Z_2 \sim \mathcal{N}(0,1)$ such that $Z_s = \rho Z_1 + \sqrt{1-\rho^2} Z_2$ and $Z_v = Z_1$.  
Following the analogy of \cite{dunn2014estimating}, we model the change in $S_t$ rather in terms of change in returns such as $\frac{S_{t+1}}{S_{t}}$; denoting the quotient by $Q_{t+1}$. 
Dividing both sides of \eqref{eq: s_disc} by $S_t$, we obtain
%
%
\begin{align}\label{eq: Q_ver}
Q_{t+dt} &=1+\mu dt   + \sqrt{v_t} \sqrt{dt}  (\rho Z_1 + \sqrt{1-\rho^2} Z_2), \\
v_{t+dt} &=v_t+ \kappa(\theta-v_t) dt   + \sigma\sqrt{v_t} \sqrt{dt} Z_1
\end{align} 
where we impose a non-negativity on variance process by defining $v_t = \max (0,v_t)$. We also remark that unlike \cite{dunn2014estimating}, we do not set $dt$ as 1 ensuring the variance of the increment over the time step is proportional to $dt$.
Recall that the likelihood function is merely the product of the probability density functions of each observations, for instance \cite{iacus2008simulation}, such that     the probability density functions at each observation point is denoted as $f(Q_{t+1}, v_{t+1})$.
To employ MLE, the joint probability density function is needed.
Given the discretization and $Z_S$, $Z_v$ are standard normal variables; inferring $Q_{t+1} \sim \mathcal{N}(1+\mu, v_t)$ and $v_{t+1} \sim \mathcal{N}(v_t + \kappa(\theta - v_t ), \sigma^2 v_t)$ and fixing $\Delta t$ as one for simplicity.
Following \cite{dunn2014estimating},
\begin{equation} \label{eq1}
	\begin{split}
		f(Q_{t+1}, v_{t+1}) & = \frac{1}{2\pi\sigma v_t \sqrt{1-\rho^2}} \exp  
		\Bigg[  
		-\frac{(Q_{t+1}-1-\mu)^2}{2v_t(1-\rho^2)} 
		\\ &+ \frac{\rho(Q_{t+1}-1-\mu)(v_{t+1}-v_t-\theta \kappa + \kappa v_t)}{V_t\sigma(1-\rho^2)} \\
		&- \frac{v_{t+1}-v_t-\theta \kappa + \kappa v_t)^2}{2\sigma^2 v_t(1-\rho^2)} 
		\Bigg].
	\end{split}
\end{equation}
For ease of computation, taking natural logarithm of the likelihood function yields given in \eqref{eq1} is implicitly given by
\begin{equation}
	 \ell (\mu, \kappa, \theta, \sigma, \rho)=\sum _{t=1}^{n} \ln (f_{Q, v}(Q_{t+1}, v_{t+1})\mid \mu, \kappa, \theta, \sigma, \rho), 
\end{equation}
for which taking the partial derivatives of $\ell ( \cdot)$ with respect to the parameter set $[\mu, \kappa, \theta, \sigma, \rho]$ yields the log-likelihood function such as
\begin{equation} \label{eq:loglikelihood}
	\begin{split}
		\ell(\mu,\kappa,\theta,\sigma,\rho) = \sum_{t=1}^{n} \Bigg[ 
		& -\ln(2\pi) - \ln(\sigma) - \ln(v_t) - \frac{1}{2}\ln(1 - \rho^2) \\
		& - \frac{(Q_{t+1} - 1 - \mu)^2}{2v_t(1 - \rho^2)} \\
		& + \frac{\rho (Q_{t+1} - 1 - \mu)(v_{t+1} - v_t - \theta \kappa + \kappa v_t)}{v_t \sigma (1 - \rho^2)} \\
		& - \frac{(v_{t+1} - v_t - \theta \kappa + \kappa v_t)^2}{2 \sigma^2 v_t (1 - \rho^2)} 
		\Bigg],
	\end{split}
\end{equation}
by which we refer to \cite{dunn2014estimating} for further technical details. 
Minimizing \eqref{eq:loglikelihood} with respect to the parameters at hand, we could easily obtain the estimates.
Now we move on to the sliced inference.

\subsection{Sliced inference}

Let $X=X_{\text{all}} \in \mathbb{R}^{n \times d} $ be our high-dimensional feature space with rows $x_i ^\top$ such as
\begin{equation} \label{matrix: X_all}
	X_{\text{all}} = \begin{bmatrix}
		X_{1}^{(1)} & X_{2}^{(1)} & \dots & X_{d}^{(1)} \\
		X_{1}^{(2)} & X_{2}^{(2)} & \dots & X_{d}^{(2)} \\
		\vdots & \vdots & \ddots & \vdots \\
		X_{1}^{(n)} & X_{2}^{(n)} & \dots & X_{d}^{(n)}
	\end{bmatrix}
\end{equation}
and let $Y\in \mathbb{R}^d$ be the target (or response) variable. Our aim, fundamentally, is to obtain a low-dimensional data (subspace) spanned by effective dimension reduction directions $\beta_1, \ldots, \beta_B \in \mathbb{R}_d$ such that
\begin{equation}
	\mathbb{E}[Y \mid X] = \mathbb{E}[Y \mid X \beta_1, \ldots, X \beta_B ],
\end{equation}
the equality above holds, and definition of $B$ is due soon.
The corresponding estimation starts with standardization of $X$, with $\bar{x}^s \in \mathbb{R}^d$ be the sample  mean vector, and  $\Sigma_X^s$ be the sample covariance.
Standardization is then $z_i = ({\Sigma_X}^s) ^{-1/2}(x_i - \bar{x}^s)$  for $ i = 1, \ldots, n  $ yielding a standardized data set $Z \in  \mathbb{R}^{n \times d}$. 
Being a supervised dimensionality reduction methods, the target variable is sliced by partitioning the range of $Y$ into $M$ disjoint slices (intervals) $H_m$, for $m=1,\ldots, M$. 
With $n_m$ denoting the number of observations within tge slice $H_m$, and $I_{H_m}$ being indicator function for this slice such that
$n_m = \sum_{i=1}^{n} I_{H_m}(Y_i)$.
As also put forward by \cite{hardle2024computationally}, if not sliced; SIR coincides with PCA. 
Compute the mean of $z_i$ over all slices $\tilde{z}_m = {1}/{n_m}\sum_{i=1}^{n} z_i I_{H_m}(Y_i)$, with slice-wise means computed; estimation of the covariance of the inverse regression functions is now possible so that 
\begin{equation}
\hat{V} = n^{-1}\sum_{m=1}^{M} n_m \tilde{z_m} \tilde{z_m}^\top
\end{equation}
Upon estimating $\hat{V}$, a critically important aspect of dimensionality reduction and being able to understand the structure of the data lies in the decomposition of  $\hat{V}$ into its eigenvalues and eigenvectors.
The decomposition is expected to provide further insight on the variability of the data.
Therefore the algorithm then computes the eigenvalues $\hat{\lambda}_i$ and eigenvectors $\hat{\eta}_i$ of $\hat{V}$, and select the top $B$ eigenvectors corresponding to the largest eigenvalues; capturing the ones explaining more variability in the data.
The reason to select the top eigenvalues is the indication that they could offer higher explanations in the  variability which could lead to better precision in our case.
Therefore now we define $W$ as $[\hat{\eta}_1 \cdots \hat{\eta}_B] \in \mathbb{R}^{d \times B}$.
However, in original SIR in \cite{li1991sliced}, there is one more step which transforms the standardized EDR directions $\hat{\eta}_i$ back to the original scale such that $\hat{\beta}_i = {\Sigma_X^s}^{-1/2} \hat{\eta}_i.$ for $i = 1, \ldots, B$.
The reason we skip the last step might seems unreasonable, yet there is a chance of numerical instability that could be amplified during the last step; especially if the sample covariance is ill-conditioned. 
But most importantly, we find no specific reason to revert back to the original scale as obtaining reduced-space representation is sufficient.
For numerical issues, and an interesting comparison of (de-)standardization approaches we refer to \cite{kessy2018optimal}.
Furthermore, given the extra computational cost we observe no significant improvement in Mean Squared Error (MSE) values, therefore we prefer to work in the standardized low-dimensional data.


We asserted that we are following \cite{dunn2014estimating} in the Heston dynamics given in \eqref{eq: Q_ver}, and its corresponding formulation of returns rather than the asset price path.
Given that variance process drives the asset price process and not the other way around, a possible selection of target variable would obviously be the asset price path, namely $S$.
However, there is a technical   reason not to do so. 
Selection of $S$ as the target variable, given the dynamics of Heston model, could 
cause unfavorable dynamics introduced to the reduced space.
Based on the Heston model, the distribution of $S$ is likely to be heavily skewed signifying a possible problem with the slices being equally width.
For instance assuming 10-slice bins across the range of the target variable; the sort would mean the target variable sliced into 10 equal width intervals in terms of its value.
What contributes to this possible issue is the likelihood of some slices having fewer or even no observations considering the possibility of heavily skewed distribution of the asset price process.
This issue could magnify especially around the end-tails of the distribution.
Therefore, our preference, we remark, to work with returns, $Q$,  is justified by transforming our target variable into less skewed and more symmetric.
Although our return structure is strictly positive as well, the original SIR imposes no non-negativity constraint on the target variable, $Y$; making it possible to work with logarithmic returns as well.
We leave this suggestion to market practitioners given the modeling requirements might vary.
We acknowledge, however, deeper perspectives require asymptotic analysis of such matters.
With that, we refer to an excellent discussion of asymptotic behavior of SIR given in \cite{hsing1992asymptotic}, along with the study in which various slicing approaches are discussed in \cite{coudret2014new}

In our set-up, the structure $X=X_{all}$ depends solely on our preference of order and inclusion; and we define the $i$-th  row as $  [ \mu, \kappa, \theta, \sigma, \rho, v_t^i  ]$ for $i = 1, \ldots, n$. 
First, we note, the inclusion of the parameters to be estimated seems highly necessary, along with the other dimensions of a system of stochastic differential equations.
Since in our case, we only have one (i.e., variance process); this becomes a higher priority.
We also remark that we have 6 elements in the row, yet this also is a flexible choice.
Some candidates might include lagged values of the variance process, or cross products such as $\kappa (\theta- v_i)$ might increase to the explanatory power of the procedure.Moreover, $X_{\text{all}}$ is a data set of predictors; meaning that if subject-related expertise suggests including a particular one, the possible effects should be thoroughly analyzed.

After applying the steps of SIR briefly laid above with  $W$ as the matrix of eigenvectors corresponding to the most significant components in the SIR transformation; we obtain
\begin{equation}
X_{\text{reduced}} = X_{\text{all}} \cdot W
\end{equation}
where  the transformation finds the low-dimensional structure in the data that best explains the variation in the target variable.
SIR does not explicitly estimate the parameters of the Heston model, rather it reduces the dimensionality of the data. 
Therefore, the log-likelihood given in \eqref{eq:loglikelihood} is no longer valid and applicable in this case.
The reason is straightforward; we no longer explicitly have paths generated but instead a matrix filled with predictors for each path.
This new approach should have a different approach.
Given we have MLE for the Heston in the previous section, we use MLE to fit  the Heston model to the reduced feature set obtained from SIR. 

The idea is actually quite simple, given $X_{\text{reduced}}$ still contains stochastic information in regard to parameters.
This leads to the possibility of minimizing $X_{\text{reduced}}$, yet now the only problem is that $X_{\text{reduced}}$ does not have any functional representation on which we could minimize to extract the parameters.
Even though we transformed the feature space $X$, the underlying structure of the data still follows a probability distribution; reducing the dimensions doesn't change the probabilistic structure
Recall that MLE does not minimize (directly) on (any) raw data. 
It minimizes the negative log-likelihood function describing the probability of observing the given data under a parameterized model.
With $X_{\text{reduced}} = X_{\text{all}} \cdot W$, we conclude that  the likelihood function can still be expressed in terms of original model parameters which gives us a function to optimize.
Therefore the mapping is still differentiable, and $X_{\text{reduced}}$ still following a probabilistic structure; MLE works mathematically.

Let us express the transformed likelihood function explicitly.
Similar to the normality assumption given in \eqref{eq1}, suppose the original data follows a multivariate normal distribution:
\begin{equation}
X_{\text{all}} = X \sim \mathcal{N}(\mu_X,~ \Sigma_X)
\end{equation}
where $\mu_X \in \mathbb{R}^d$ and $\Sigma_X \in \mathbb{R}^{d \times d}$ are functions of $\mu$, $\kappa$, $\theta$, $\sigma$, $\rho$.
Since $X_{\text{reduced}}$ is a linear transformation of $X_{\text{all}}$, we obtain
\begin{equation}
X_{\text{reduced}} = X_{\text{all}}   W \sim \mathcal{N}(\mu_X W,~ W^\top \Sigma_X W)
\end{equation}
where $W^\top \Sigma_X W \in \mathbb{R}^{d \times d}$ and $\mu_x W  \in \mathbb{R}^B$. 
Given our multivariate normality assumption, the derivation of the log-likelihood of the reduced data becomes trivial.
We also like to remark that a major advantage of our contribution is the generality of the log-likelihood of the reduced data. 
There is no need to derive a specific log-likelihood function for each SDE, the following log-likelihood is applicable as long as the normality assumption is not too deviated from the underlying modeling principles. 
With $n$ number of observations (number of paths in our case), $B$ is 6 and setting $Z = (W^\top \Sigma_X W) $; the likelihood function is simply found as
\begin{equation}
\ell (\cdot)= \prod_{i=1}^{n} \frac{1}{(2\pi)^{B/2} \, \det(Z)^{1/2}} \exp  
\Bigg[  
-\frac{1}{2} (X_{\text{reduced}, i}-
 \mu_X W)^\top (Z)^{-1} (X_{\text{reduced}, i} - \mu_X W), \notag
\Bigg]
\end{equation}
and taking natural logarithm of both sides simply yields
%
%
%
%
%
\begin{equation}\label{log_reduced}
\begin{split}
	=&\frac{nB}{2}\ln(2\pi)-\frac{n}{2}\ln\det(W^\top\Sigma_X W) \\
&-\frac{1}{2}\sum_{i=1}^n(X_{\mathrm{reduced,}i}-W^\top\mu_X)^\top(W^\top\Sigma_XW)^{-1}(X_{\mathrm{reduced,}i}-W^\top\mu_X),
\end{split}
\end{equation}
with $\mu_x \in \mathbb{R}^d$,  $\mu_x W  \in \mathbb{R}^B$, $X \in \mathbb{R}^{n \times d}$,  $W^\top \Sigma_X W \in \mathbb{R}^{d \times d}$, and the projection matrix with $W  \in \mathbb{R}^{d \times B}$.
In implementation-wise, we feed MLE with $X_{\text{reduced}}$ and $W$.
We now focus on the derivation of $\mu_X$ and $\Sigma_X$.

Another major advantage of assuming $Q_t$ as simple returns is on approximating the expectation and variance of $Q_t$. 
Given $Q_{t + \Delta t}$ is equal to $1 +\mu \Delta t+ \sqrt{v_t \Delta t}Z_s$, the expectation is, trivially, found as $1+\mu \Delta  t$, and the variance is obtained similarly, $\mathbb{V}(Q_t)$ is approximated by $\mathbb{V}(\sqrt{v_t \Delta t} Z_s)$; which could be written as
\begin{align}
\mathbb{V}(Q_t) &\approx \Delta t \bigg( \mathbb{E}(\sqrt{v_t})^2 \mathbb{V}(Z_s) + \mathbb{E}(Z_s)^2 \mathbb{V}(\sqrt{v_t})  + \mathbb{V}(\sqrt{v_t}) \mathbb{V}(Z_s)       \bigg ) \\
&\approx \Delta t \bigg( \mathbb{E}(\sqrt{v_t})^2 + \mathbb{V}(\sqrt{v_t})         \bigg)
\end{align}
where $\mathbb{E}(\sqrt{v_t})^2$ is equal to $\mathbb{E}(v_t) - \mathbb{V}(\sqrt{v_t})$ which leads to $\mathbb{V}(Q_t)$ being approximated by  $\mathbb{E}(v_t) \Delta t$.
From \cite{rouah2013heston}, we know $\mathbb{E}(v_t)$ is given by $\theta + (v_0-\theta) \exp(-\kappa t)$, yet assuming stationarity for the variance process as $t \rightarrow \infty$, expectation of the variance process $\{v_t\}^T_{t=0}$ becomes $\theta$ and $  \mathbb{V}  (Q_t) $ is approximated by $\theta \Delta t$.  
Assuming stationarity of the   variance process $\{v_t\}^T_{t=0}$, $\mathbb{V}(v_t)$ could be well approximated by $\sigma^2 \theta / 2\kappa$.
This is where we differ from \cite{dunn2014estimating} as we approach the variance process a bit more asymptotically.
Yet we remark if we are to follow \cite{dunn2014estimating} in that regard as well;   $\mathbb{V}(v_t)$ could be well approximated by $\sigma^2 \theta \Delta t$.
We also need to explain why we are not multiplying by $\Delta t$. 
$\sigma^2 \theta / 2\kappa$ comes from the continuous-time Heston model at the steady-state, and there is no need to rescale it with $\Delta t$; this holds for the following discussion as well.

To complete the covariance matrix; $\mathbb{C}(Q_t, v_t)$ and $\mathbb{C}(v_t, Q_t)$ are required.
Given the symmetric nature, deriving one would suffice.
Although deriving the covariance matrix of a multidimensional SDE is  meticulous, the procedure is straightforward; an excellent example could be found in \cite{grzelak2011heston}.  
However, we advocate a surrogate one with enough parameterization that is comprehensive, but simple enough so as not to increase the complexity of minimization during MLE.
Given the correlation, $\rho$, encoding the dependence between $\{Q_t\}^T_{t=0}$ and $\{v_t\}^T_{t=0}$, we suggest that the product of $\rho$ and $\mathbb{V}(v_t)$ could approximate the covariance. 
To simplify, $\mathbb{V}(v_t)$ measures the variation in the variance process, and $\rho$, through 
$\rho \mathbb{V}(v_t)$, projects the variation in the variance process onto the direction of $\{Q_t\}^T_{t=0}$, and approximating somewhat of a joint structure. 
Hence, $\rho \mathbb{V}(v_t)$ could be assumed as (approximately bivariate) covariance.
Although this approximation requires approximately linear relationship between $\{Q_t\}^T_{t=0}$ and $\{v_t\}^T_{t=0}$, considering the possible loss of information after dimension reduction via \emph{SIR}, we remark our proxy as plausible.

So far our calculations imply that the mean vector, $\mu_X$ has 2 elements and  $\Sigma_X$ is a $2 \times 2$ matrix. 
However, $X$ is, by construction, size of $d \times d $, leading to $\mu_X \in \mathbb{R}^d$ and $\Sigma_X \in \mathbb{R}^{d \times d}$, similarly by definition.
Through our selection $X$, we require no mismatch in further calculations. 
Therefore, already having 2 elements (expectations) for $\{Q_t\}^T_{t=0}$ and $\{v_t\}^T_{t=0}$, we assume dummy values and recall $\Delta t$ is taken one such that
\begin{equation}
	\mu_X = [ (1+ \mu ) \quad \theta \quad 0 \quad 0 \quad 0 \quad 0]^\top,
\end{equation}
since there are no other processes to learn from besides the price and variance processes, assigning zeros as a proxy value makes sense considering the first two captures all the necessary information. 
We also remark that assigning very small numbers instead of zero is also plausible as during our tests we observe no discernible improvement in Mean Squared Error values.

Similar assumption is also required for the covariance matrix of $X$ given $\Sigma_X \in \mathbb{R}^{d \times d}$.  
As the fitted parameter space is simply $\{\mu, \kappa, \theta, \sigma, \rho \}$, recall that $d$ is not the number of parameters to be estimated, but the number of columns (dimensions) of $X = X_{all}$ in \eqref{matrix: X_all}.

Given d is 6, and we have variances and symmetric covariances of $\{Q_t\}^T_{t=0}$ and $\{v_t\}^T_{t=0}$ we order the columns of the following matrix such as $\{Q, v, \mu, \kappa, \theta,  \rho \}$; considering the inclusion of $\rho$ might carry higher importance than that of $\sigma$ and $\mu$ is the only parameter in $\{Q_t\}^T_{t=0}$; we omit volatility of volatility parameter. However, building on the idea of   the  uncorrelated parameters, the ordering or the inclusion into $\Sigma_X$ of a particular parameters  stands irrelevant.

\begin{equation} \label{matrix:covariance}
	\Sigma_X=
	\begin{bmatrix}
		\theta  & \frac{\rho\sigma^2\theta}{2\kappa} & 0 & 0 & 0 & 0 \\  
		\frac{\rho\sigma^2\theta}{2\kappa} & \frac{\sigma^2\theta}{2\kappa} & 0 & 0 & 0 & 0   \\ 
		0 & 0 & \delta & 0 & 0 & 0 \\
		0 & 0 & 0 & \delta & 0 & 0 \\
		0 & 0 & 0 & 0 & \delta & 0 \\
		0 & 0 & 0 & 0 & 0 & \delta,
\end{bmatrix}
\end{equation} 
where elements diagonally refer to the variances.
While acknowledging $\{\mu, \kappa, \theta, \sigma, \rho \}$ constant, thus admitting to no variability, in order not to have any stability issues in our calculations, we assume $\delta$ is a sufficiently small number, in our illustrations we set $\delta$ as $0.000001$. 
Off-diagonal values, on the other hand, call for reason.
In our work, we assume no correlation between the parameters, as there are already fixed values (i.e., constants) indicating correlation among the parameters is not a question to seek answers for.
In practice, however, such assumptions might be unrealistic depending on the perspectives of the model requirements.
For instance, faster mean reversion might indicate lower $\theta$, thus interposing possible correlation among $\kappa$ and $\theta$.
In matters of calibration, especially with time-dependent parameters such as $\mu(t)$ or $\sigma(t)$, this issue might need immediate attention.
Yet, in our approach, we assume the structure presented in \eqref{matrix:covariance}. 
 
Though we impose the necessity of such   assumptions given the dimensional structure of $X$ and what $X$ is reduced to ($X_{\text{reduced}}$), we remark that the additional values (i.e. zeros) in $\Sigma_X$ and $\mu_X$ along with their interpretations might require expert knowledge on the specific model at hand.
We argue our assumption in regard to the other elements in both is plausible, yet further research is needed.
Therefore, we leave this modeling perspective as future research.

\section{Numerical Illustrations}

In\footnote{All scripts are in Python and all associated numerical illustrations presented in this manuscript are carried out on a system with i7 Core with 2.20 GHz and 16 GB RAM.} optimizing either \eqref{eq:loglikelihood} or \eqref{log_reduced}; several issues should be considered in advance.
Although convexity is desired as it relieves the modeler from the concerns of local minima.
However, with financial perspectives, especially on using MLE\footnote{Joint estimation across all paths; here (Global) MLE aims to estimate a single set of parameters that fits the entire data set of paths. This is done by maximizing the total likelihood across all paths simultaneously; and we refer to this \emph{Direct Inference}.} for models like the Heston model directly or indirectly (i.e. reduced); the log-likelihood (the objective function in our case) might not always be a reasonable expectation.
The reason is straight forward, the increased possibility of multiple local minima due to jumps or the existence of stochastic volatility etc.
Particularly with the Heston model, parameters space including mean reversion speed, correlation or volatility could induce highly complex and non-convex surfaces. 
Although the log-likelihood function is likely to be differentiable almost everywhere, assuming no issues on invertibility in \eqref{log_reduced}; we recommend caution in applying our suggested method of inference in cases of discrete jumps in the asset price path.
Supposing that even if the objective function is differentiable, recall the SIR transforms the data at hand; we note that this transformation might introduce noisy behaviors for which we are not aware of any studies conducted on such matters.

Another source of issue might be the smoothness of the objective function.
In case of financial models introducing sharp changes, spikes or even the stochastic volatility component itself might lead to no smoothness.
Without further and deeper research we find it difficult to be precise; however, the transformed or reduced features are likely to be smooth given that they are rather based on the eigenvectors of the covariance matrix of $X_{\text{all}}$.
This aspect might also be taken into consideration as \eqref{log_reduced} might not be a smooth function.
Therefore; in our study we exercise great caution in initial values of the parameters as well as with the bounds of the parameter space.
Another reason to do so includes the Feller condition.

Considering that we find it highly useful to enforce constraints on parameters and being in high dimension of parameter space, L-BFGS-B (Limited-memory Broyden-Fletcher-Goldfarb-Shanno with Box Constraints), a Quasi-Newton method efficient for large-scale optimization problems could be reasonable choice of optimization algorithm.
Given that it is sensitive to local minima; we also embed economic interpretation of parameter values and possible.
For its efficiency in convergence, the algorithm could tend well to smooth and differentiable objective functions.
Yet we remark that it would be   interesting  to carry out a comparison in sorts for which we leave open for the time being.

Given that we prefer to operate under the real-world measure $\mathbb{P}$ rather than $\mathbb{Q}$; we have drift term rather than the interest rate (risk-free rate) as the input. 
The drift, theoretically, defines the long-term expected return of the asset.
In our study we set the true value of $\mu$ at $0.03$; so as not to have the asset price overcome by the drift completely.
We find $\mu = 0.03$ reasonable as it reflects a conservative estimate (or expectation) aligned with the findings of a famous study given in \cite{fama2002equity}.

Although it's hard to impose an upper limit on $\kappa$, we set the true value of the mean-reversion speed at 5 following the discussions in \cite{rouah2013heston}.
We also justify our choice suggesting that lower values of $\kappa$ could indicate the volatility persisting for longer periods.  
If   $\kappa$ is too small, the volatility process might also be too prone to large deviations.
An interesting aspect of the Heston model is the inclusion of long-term variance, $\theta$.
A lower value of $\theta$ might indicate that the underlying asset could tend to be less volatile.
For reasons of increased difficulty in fitting parameters, we do not aim to allow such setting.
Therefore, the choice of $\theta=0.05$, we expect, could indicate realistic market behavior.
In conjuncture with $\theta$; the volatility of the volatility plays an important role capturing the degree of fluctuation.
Similarly we follow our thinking of the necessity of modesty in simulated market conditions, we set $\sigma = 0.2$.
Besides; $\sigma \sqrt{v_t}$ might cause large fluctuations, especially if the initial value of variance process, $v_0$ is considerably large.
This might lead to explosive or erratic paths.
Therefore we set $S_0 = 10$ and $v_0 = 0.01$.
Also the value of the volatility of volatility might seem quite large; yet if we assume there are 250 time steps (i.e. trading days in a year); $\sigma_{\text{daily}} = \sigma \times \sqrt{1/250}$ which approximately delivers a 1 per cent on a daily basis.

The correlation between the two-dimensions of the Heston model is an interesting feature.
For no reason  other than curiosity we induce negative correlation between the asset price process and the variance process, that is, $\rho = -0.5$.
Besides, negative correlation could be observed as the fall of prices and the increase of the volatility are jointly quite reasonable.
To ensure the Feller condition is not violated; we also impose upper and lower boundaries on our parameter space.
Therefore; $\mu \in  [-0.05, 0.05]$, $\kappa \in [5, \infty)$, $\theta \in [0.01, 0.05]$, $\sigma \in [0.00001, 0.3]$, and finally $\rho \in (-0.9, 0.7)$.
Although in some cases the bounds might seem restrictive; without such bounds there exist a likelihood of negative variance values given the Feller condition not being met.

We start with $n=1$ (i.e. one path) as real-life data dictates such harsh reality. 
Although in the later part we will carry out direct inference (DI) and sliced inference (SI) on multiple paths generated by the Heston model with true parameters outlined above; we find it reasonable to focus on the comparison of such a case.
We calculate our metric of performance based on the values we obtain along the path; rather than focusing on the final value, $S_T$.
A major reason is our interest in observing the capability and the predictive power of sliced inference on all points in time.
Throughout this section, we have 10 slices and we set $B=6$. 
Besides; we assume there are 250 days of trading in a year with $T$ as a year; such that
\begin{equation}
\text{MSE} = \frac{1}{n \cdot T} \sum_{i=1}^{n} \sum_{t=1}^{T} \left( S_{\text{est}}^{(i)}(t) - S_{\text{true}}^{(i)}(t) \right)^2
\end{equation}
Where $n$ is the total number of paths, $T$ is the number of time steps in each path, $S_{\text{est}}^{(i)}(t)$ is the predicted price at time step $t$ for path $i$, based on the fitted parameters and $S_{\text{true}}^{(i)}(t)$ is the true price at time step $t$ for path $i$ for $i = 1, \ldots, n$, based on the simulation.
It sums the squared differences between the estimated and true prices for each time step of each path, then averages over all paths and time steps.
Yet we remark that practitioners focused on derivative pricing might prefer to deal with final prices generated by the asset price path.
This metric is closer to path-dependent option pricing approach in terms of its expectations.
The initial guess to start the optimization procedure we select the initial values as $[0.02,\quad 5.2,\quad 0.04,\quad 0.1,\quad -0.6]$.
Given that working with one path might be unfavorable, noisy or even unrepresentative to present the potential of sliced inference; we simulate a single path based on different seeds for randomness \footnote{We prefer np.random.seed() of Numpy's for which we randomly select 15 seeds from 1 to 1000.} up to 15 runs of experiment.
Singular path here refers to the first and only the first row of \eqref{matrix: X_all}.
\begin{table}[htbp]
	\centering
	\caption{Results of single path}
	\begin{tabular}{ccc}
		Number of Simulations & MSE-DI & MSE-SI   \\
		\midrule
		10    & 0.6294764 & 0.248428 \\
		15    & 0.7869194 & 0.280872 \\
	\end{tabular}%
	\label{tab:1 }%
\end{table}%

\begin{figure}[h]
	\centering
	\includegraphics[width=0.8\linewidth]{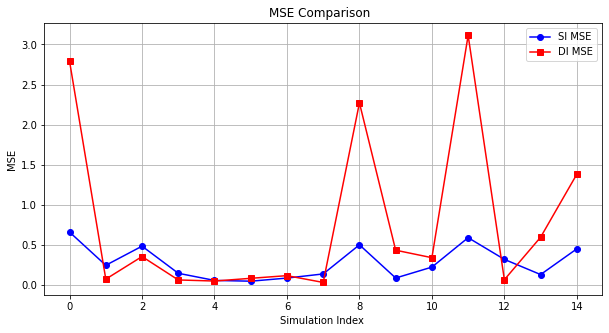}
	\caption{ MSE comparison for single paths with different sources of randomness}
	\label{fig:mse1}
\end{figure}

We observe the results of performance of both methods of inference in Table \ref{tab:1 }.
Interestingly and unexpectedly, Sliced Inference outperforms Direct Inference up to 15 different single paths with different fixed randomness.
The reason we do not increase the number of simulations is that we continue to observe such results.
Although we operate under randomness, and various randomized paths might lead to different results of performance; Sliced Inference, we observe, offers an interesting finding for which we now refer to Figure \ref{fig:mse1}.

Figure \ref{fig:mse1} shows an interesting feature of Sliced Inference, signaling that our suggested approach might be a strong candidate even as a first method of estimation.
When estimating on single paths, DI  produced extreme outliers in MSE, indicating high sensitivity to path-specific noise. 
In contrast, SI demonstrated more stable performance, with significantly fewer extreme deviations.
This could suggest that SI may offer better robustness under limited data scenarios or in the presence of high stochastic variability.
Although in Figure \ref{fig:mse1} we observe lower values of MSE of DI occasionally (hence indicating possibly better performance as $n$ increases).
We conclude that \emph{Sliced Inference} might be a better alternative on-single path cases as it seems to edge off the extreme deviations likely due to the limited data at hand.
Given that increased number of samples is likely to efficiency of \emph{Direct Inference} for statistical reasons we do not discuss here; we now present a similar version of Figure \ref{fig:mse1} with 50 paths, yet generated on the same seed of the random number generator. 
\begin{figure}[h]
	\centering
	\begin{subfigure}{0.6\textwidth}
		\centering
		\includegraphics[width=\linewidth]{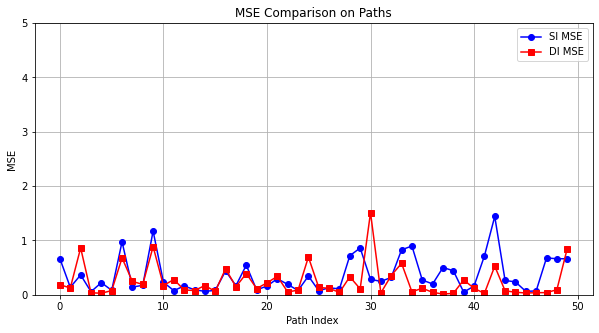}
		\caption{Without extreme outlier}
		\label{fig:d}
	\end{subfigure}
	\hfill
	\begin{subfigure}{0.6\textwidth}
		\centering
		\includegraphics[width=\linewidth]{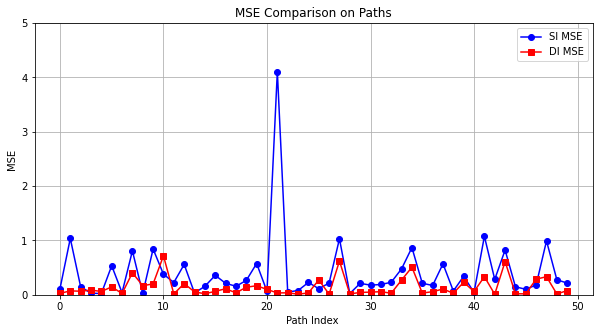}
		\caption{With extreme outlier}
		\label{2:d}
	\end{subfigure}
	\caption{MSE comparison for multiple paths with fixed source of randomness}
	\label{2}
\end{figure}
In \ref{2} we observe comparison of MSE with and without extreme outliers. 
We note that with increased number of available data \emph{Direct Inference} (i.e. MLE) no longer generates extreme fits.
This attributes that the parameters found in accordance with the higher amount of data have a better convergence to the true parameters.
However, in both Figures \ref{fig:d} and \ref{2:d}, excluding outliers, we observe closer performance to each other.
For instance, Figure \ref{2:d} is based on the results provided in the first row of Table \ref{tab:addlabel}.
While \emph{Direct Inference} yields an MSE of 0.14, \emph{Sliced Inference} yields an MSE of 0.40, approximately in both cased.
Therefore, we conclude that excluding such outliers; \emph{Sliced Inference} delivers acceptable parameter estimation results.
In a way, MLE could operate on the reduced space of $X_{\text{all}}$

\begin{table}[htbp]
	\centering
	\caption{Results of   multiple paths}
	\begin{tabular}{ccc}
		$n$     & Computation Time & Average MSE \\
		\midrule
		\multirow{2}[2]{*}{50} & 47.2  & 0.142019 \\
		& 8.1   & 0.401661 \\
		\midrule
		\multirow{2}[2]{*}{100} & 90.7  & 0.175657 \\
		& 15.3  & 0.388457 \\
		\midrule
		\multirow{2}[2]{*}{250} & 244.5 & 0.207752 \\
		& 36.3  & 0.479181 \\
		\bottomrule
	\end{tabular}%
	\label{tab:addlabel}%
\end{table}%

\begin{figure}[htbp]  
	\begin{subfigure}[c]{0.9\textwidth}
		\centering
		\includegraphics[width=0.8\linewidth]{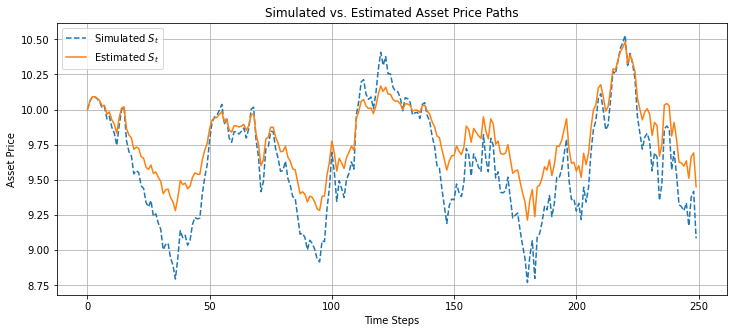}
		\caption{Direct inference}
		\label{fig:comp1}
	\end{subfigure}
	\begin{subfigure}[c]{0.9\textwidth}
		\centering
		\includegraphics[width=0.8\linewidth]{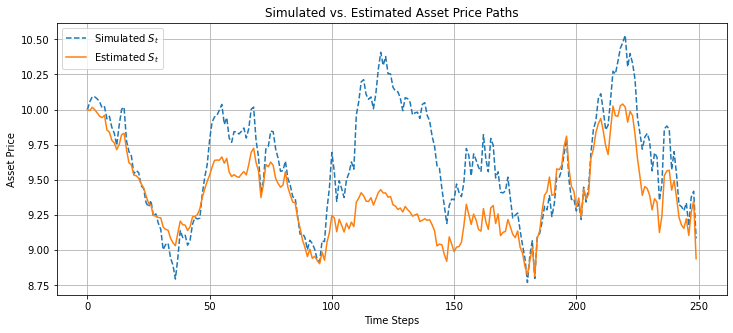}
		\caption{Sliced inference}
		\label{fig:comp1*}
	\end{subfigure}
	\caption{One of the best case of DI and one of the worst case of SI }
	\label{ex}
\end{figure}
In Table \ref{tab:addlabel}; we generate paths up to 250, and report both MSE values and the time required for computation given in terms of seconds.
Another important aspect of \emph{Sliced Inference} is the computational time required; the difference between \emph{Sliced Inference} and \emph{Direct Inference} strikes far better than expected.
Especially in cases with distinctively large number of paths simulated; such an aspect might be sought after.
Another interesting plot is a comparison of better and worse performance of both SI and DI.
In Figure \ref{ex}; we observe such examples purposefully selected out of simulated paths.
Although the trend is followed closely, path-wise fitting is highly under-performing.

\begin{figure}[h]  
	\begin{subfigure}[c]{0.5\textwidth}
		\centering
		\includegraphics[width=0.8\linewidth]{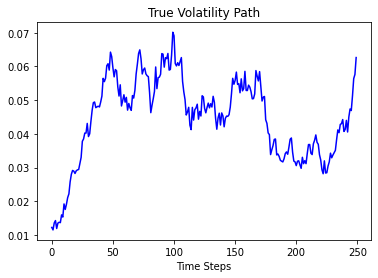}
		\caption{Direct inference}
		\label{1}
	\end{subfigure}
	\begin{subfigure}[c]{0.5\textwidth}
		\centering
		\includegraphics[width=0.8\linewidth]{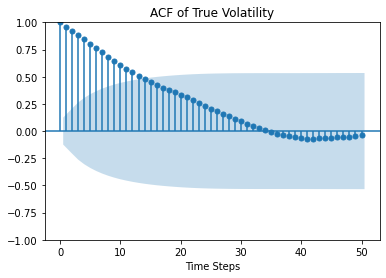}
		\caption{Direct inference}
		\label{2*}
	\end{subfigure}
	\begin{subfigure}[c]{0.5\textwidth}
		\centering
		\includegraphics[width=0.8\linewidth]{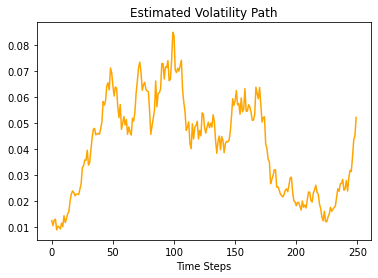}
		\caption{Sliced inference}
		\label{3}
	\end{subfigure}
	\begin{subfigure}[c]{0.5\textwidth}
		\centering
		\includegraphics[width=0.8\linewidth]{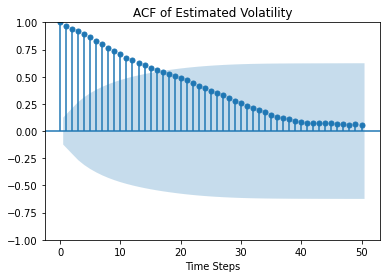}
		\caption{Sliced inference}
		\label{4*}
	\end{subfigure}

	\caption{Volatility paths and ACFs of a randomly selected path }
	\label{var1}
\end{figure}

Though we so far focused on the price process, the Heston model also incorporates a variance process.
Therefore; brief and visual analyses of \emph{Direct Inference} and \emph{Sliced Inference} are timely.
In Figure \ref{var1} we observe  true and estimated volatility paths and the corresponding  autocorrelation functions of a randomly selected path with fixed randomness for each case.
Given a characteristic feature of the Heston model, expecting volatility clustering is reasonable and ensured in Figures \ref{2*} and \ref{4*}.

Although there are other aspects of the Heston model and views of statistical requirements in matters of estimation that might also be addressed.
However, we believe, to be concise, our arguments should suffice to remark that \emph{Sliced Inference} could be accepted as a reasonable approach.
We, again on purpose, neither carry out sensitivity analysis of the fitted parameters nor report any performance on such manner; simply because we are more interested in path trajectories. 
We close this section with final remark that the underlying approach of \emph{Sliced Inference} (i.e. the SIR) is a model-free approach.
Therefore; this reality and flexibility of \emph{Sliced Inference} should always be considered in further implementations possibly by market practitioners or academicians.

With this our contribution has several aspects worth mentioning.
Firstly, our proposed approach does not explicitly require the derivation of log-likelihood function for each case; as it is a general case that could easily be adapted to any model with much simpler steps.
Its model-free structure allows to work with any sort of data, and frees the modeler in a practical way.
The results obtained on single paths show that \emph{Sliced Inference} could be safely used as it answers quite well to the limited data and lack of asymptotics.
We also remark that in cases of large amount of simulated data; our approach offers significance gains in terms of computational time, while delivering acceptable predictive performance.
We also note that  in our case of modeling; \eqref{matrix: X_all} does not have variability in the first 5 columns.
Inclusion of variable components such as lagged variables or identities like $\sigma S_t$ might even increase the predictive power of \emph{Sliced Inference}.

\section{Conclusions}

Based on our findings; we firmly believe that our suggested method of estimation of parameters in multi-dimensional stochastic differential equations performs quite reasonably.
Surely, there is loss of information expected in implementation-wise; yet \emph{SIR} could be a valid and dependable dimensionality reduction method on models such as the Heston model.

We humbly assert the need to seek after methods requiring less data rather than more.
Therefore; we hopefully believe that we fulfill our purpose in presenting sort of such methods structurally.
Although we stress again the nature of the \emph{SIR} being \emph{model-free}; the flexibility overall provided is fundamental in many financial applications where analytically tractable methods might be costly or even absent.
Introducing stochastic interest rates, for instance, to the Heston model not only increases the dimension of SDE to three but also increases the challenge of estimation; \cite{grzelak2011heston}.
So that we argue our suggested approach could handle even possible extensions on already challenging models.

In structuring the features, $X=X_{\text{all}}$; we followed the row by row structure to stack all the components from each path separately.
This matrix then has one row if there is a single path (or historical data) or up to $n$ rows if there are $n$ paths.
However; we believe this could be highly useful in estimating time-dependent parameters of the Heston model or any other.
For instance; with one path but $T$ amount of $\mu(t)$ for $t \in T$, there then becomes a stack of $\mu(1), \mu(2), \ldots, \mu(T)$ on each row.
Given the difficulty of estimating time-dependent parameters via MLE or some other way; \emph{Sliced Inference} could bring in significant ease and value in retrieving such time-dependent parameters.  
We therefore leave this open for now as a future research direction.
Similarly; with term structure models an approach with $X=X_{\text{all}}$ might be highly beneficial.
In a further study; we aim to address such modeling perspectives.

\backmatter


\section*{Compliance with Ethical Standards}


\bmhead*{Competing Interests}
 There are no financial or non-financial interests   directly or indirectly related to the work submitted for publication.



\bigskip

\begin{appendices}






\end{appendices}


\bibliography{MyBiblio}
\end{document}